\documentclass[twocolumn,showkeywords,aps,ifacmtg]{revtex4}
\usepackage{amssymb}
\usepackage{graphicx}
\usepackage{times}
\usepackage{dcolumn}

\def\scoc {Sr$_2$CuO$_2$Cl$_2$}
\def\lco {La$_2$CuO$_4$}

\begin{document}

\title{Measurement of magnetic excitations in the two-dimensional
    antiferromagnetic \scoc~ insulator using resonant x-ray
    scattering: Evidence for extended interactions
}

\author{M. Guarise$^1$}
\author{B. Dalla Piazza$^1$}
\author{M. Moretti Sala$^2$}
\author{G. Ghiringhelli$^2$}
\author{L. Braicovich$^2$}
\author{H. Berger$^1$}
\author{J.N. Hancock$^3$}
\author{D. van der Marel$^3$}
\author{T. Schmitt$^4$}
\author{V.N. Strocov$^4$}
\author{L.J.P. Ament$^5$}
\author{J. van den Brink$^{5}$}
\author{P.-H. Lin$^1$}
\author{P. Xu$^1$}
\author{H. M. R\o{}nnow$^1$}
\author{M. Grioni$^1$}

\affiliation{$^1$Institute of Condensed Matter Physics, Ecole Polytechnique F\'ed\'erale de Lausanne (EPFL),  CH-1015, Switzerland}

\affiliation{$^2$CNR/INFM Coherentia and Soft, Dipartimento di Fisica, Politecnico di Milano, I-20133 Milano, Italy}

\affiliation{$^3$Institut de Physique de la Mati\`ere Condens\'ee,
24 quai Ernest-Ansermet, Universit\'e de Gen\`eve, CH-1211, Switzerland }

\affiliation{$^4$Swiss Light Source, Paul Scherrer Institut, CH-5232 Villigen PSI, Switzerland }

\affiliation{$^5$Intitute-Lorentz for Theoretical Physics, Universiteit Leiden, NL-2300 RA Leiden, The Netherlands}

\date{\today}
\pacs{}

\begin{abstract}

We measured the momentum dependence of magnetic excitations in the model spin-1/2 2D antiferromagnetic insulator \scoc.
We identify a single-spin-wave feature and a multi-magnon continuum, with different polarization dependences. The spin-waves display a large (70 meV) dispersion between the zone-boundary points ($\pi$,0) and ($\pi$/2,$\pi$/2). Employing an extended $t$-$t'$-$t''$-$U$ one-band Hubbard model, we find significant electronic hopping beyond nearest-neighbor Cu ions, indicative of extended magnetic interactions.
The spectral line shape at ($\pi$,0) indicates sizeable quantum effects in SCOC and probably more generally in the cuprates.

\end{abstract}

\maketitle

Magnetism in low-dimensional cuprates remains of considerable interest, in relation both to the fundamental quest to understand strong electron correlation and quantum spin effects in Mott insulators, and to the search for the mechanism of high-T$_c$ superconductivity.
To lowest order, the undoped cuprate superconductors can be described by the canonical spin $1/2$ two-dimensional (2D) square lattice nearest-neighbor (NN) Heisenberg antiferromagnet, which is among the simplest and most studied models in magnetism \cite{Manousakis1991}.
The ground state displays weak classical order, reduced by quantum fluctuations at zero temperature and destroyed by thermal fluctuations at finite temperature. A possible cross-over between renormalized classical \cite{ChakravartyHalperinNelson} and quantum critical \cite{ChubukovSachdevYe} scaling, was tested experimentally in the undoped cuprates \scoc{} (SCOC) \cite{Greven1995} and \lco{} (LCO) \cite{Kastner}, and in the organometallic salt Cu(DCOO)$_{2} \cdot$4D$_2$O (CFDT) \cite{Ronnow1999}. However, while the latter shows only nearest neighbor coupling,
high-energy inelastic neutron scattering (INS) data on LCO \cite{Coldea} suggest that further-neighbor magnetic interactions influence the above mentioned scaling measurements. Fustrated further-neighbor interactions could also bring the undoped cuprates closer to the valence bond liquid proposed as mechanism for superconductivity \cite{AndersonScience1987}. 

It is therefore timely to investigate the excitation spectrum of SCOC, as an important model system. Inelastic neutron scattering (INS) measurements of SCOC have been limited to low energies and small momenta around the ordering wave vector \cite{Greven1995}. In this Letter we report the full magnetic excitation spectrum measured by resonant inelastic x-ray scattering (RIXS). We  discover a surprisingly large dispersion along the magnetic Brillouin zone (MBZ) boundary.
An analysis of the data in terms of an extended Hubbard model yields a quantitative estimation of sizeable further-neighbor electronic hopping. The resulting series of longer-ranged magnetic interactions enhance quantum fluctuations, in agreement with the reduced ordered moment. The importance of quantum fluctuations is further
revealed by differences in the spectral line shapes at the (-$\pi$,0) and (-$\pi$/2,$\pi$/2) MZB points.

SCOC is an insulating single-layer parent compound of the high-T$_c$ superconducting (SC) materials. It is isostructural to the high-temperature tetragonal phase of LCO, with La replaced by Sr, and apical oxygens replaced by Cl. The distance between adjacent CuO$_2$ planes is 18\% larger than in LCO.  AFM order develops below T$_N$=256 K with reduced (0.34 $\mu_B$) moments aligned along the (110) direction in the CuO$_2$ plane. Both the in-plane (XY) anisotropy and the interlayer coupling are very small, and SCOC is an almost ideal realization of an $S=1/2$ 2D square-lattice Heisenberg AFM \cite{Kastner}.
Magnetic excitations in SCOC have been identified by optical spectroscopies. A structure at 0.35 eV in absorption \cite{Perkins} is interpreted as a quasi-bound state of two magnons assisted by a phonon with momentum Q$_{ph}\sim$ ($\pi$,0) \cite{Lorenzana}. Two-magnon excitations at E$_{2M}$=0.35 eV in the Raman spectra suggest a superexchange energy $J\sim0.13$~eV \cite{Tokura,Blumberg}. Neither Raman nor optics are sensitive to single magnons, which are optically-forbidden spin-flip ($\Delta S=1$) excitations. 

RIXS in transition metal (TM) oxides probes excitations with mixed charge and spin character across or inside the Mott-gap \cite{Kotani,Hasan,Ghiringhelli}. At the TM L$_{2,3}$ ($2p \rightarrow 3d$) edges the large $2p$ spin-orbit interaction couples the angular momentum of the photon to the electron spin, and pure spin-flip excitations are possible \cite{DeGroot,GhiringhelliNiO}.
In the cuprates, dispersive $\Delta S=0$ excitations have been observed in LCO both at the Cu K ($1s$) \cite{Hill} and L$_3$  \cite{Braicovich} edges, and in the ladder compound Sr$_{14}$Cu$_{24}$O$_{41}$ \cite{Schlappa}. More recent theoretical and experimental work has shown that L$_{2,3}$ edge RIXS can be used to map the dispersion of  single magnons in 2D cuprates  \cite{Ament}. Therefore x-rays are an interesting alternative to neutrons, with the advantage that RIXS can be performed also on sub-mm$^3$ samples.

\begin{figure}[t!]
\includegraphics[width=3.3in]{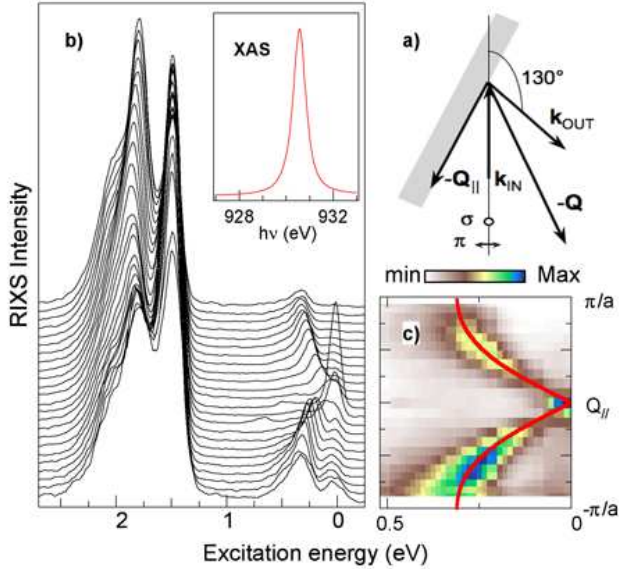}
\caption{(color online) (a) Schematics of the scattering geometry. (b) Cu L$_3$ RIXS spectra (T=15 K) along the (100) direction. The incident energy is set at the maximum of the absorption (XAS) (inset). (c) Intensity map extracted from (b). The red line is the spin-wave dispersion for the NN Heisenberg model and $J$=130 meV.
 } \label{fig:1}
\end{figure}

Measurements were performed at the SAXES end station of the ADRESS beam line of the Swiss Light Source (SLS) \cite{Saxes}. Single crystals (4$\times$4$\times$0.5 mm$^3$) grown from the flux with the $c$-axis perpendicular to the large surface, and characterized by x-ray and neutron diffraction, were mounted on a flow cryostat, with the $c$-axis and either the (100) or the (110) directions in the horizontal scattering plane. By adjusting the undulator, data were taken with incoming polarization either perpendicular ($\sigma$) or within ($\pi$) this plane. At the fixed scattering angle of 130$^{\circ}$ the transferred momentum was $\|\bf{Q}\|$=0.85 \AA$^{-1}$.
By a rotation around a vertical axis, the projection Q$_{\|}$ of $\bf{Q}$ on the $ab$ plane was varied in the range $\pm$ 0.73 \AA$^{-1}$ ($\pm$ 0.92 $\pi$/a). Positive (negative) Q values correspond to a grazing emission (incidence) geometry. The combined energy resolution was $\Delta$E=130 meV, and the accuracy on the energy zero was $\pm$ 7 meV, as determined from the elastic peak measured on a co-planar polycrystalline carbon sample. The momentum resolution, determined by the detector size, was better than $\pm$2.5$\times$10$^{-3}$($\pi/a$). Post-cleaving in situ or in air produced very similar results, confirming the good bulk sensitivity of L-edge RIXS.

Figure 1b is an overview of the spectra for the (100) direction for $\pi$-polarization They were normalized to the same integrated intensity in the 1-2.5 eV energy range, to remove intensity variations due to the angular dependence of absorption, and any other angular or time dependence. We verified that self-absorption, i.e. the partial re-absorption of the scattered beam, does not affect the determination of excitation energies over the whole range of Fig. 1. The main feature at $\sim$1.5 eV is the manifold of optically-forbidden $dd$ electron-hole excitations \cite{Kuiper,Ghiringhelli}. In the 0-1.5 eV energy range, where no electronic excitations are expected, the spectra exhibit a loss feature dispersing symmetrically from Q$_{\|}$=0 (Fig. 1c). Near the zone boundary, spectral weight extends beyond 300 meV, well above the highest phonon mode (70 meV) in SCOC \cite{Zibold}. Its maximum follows the calculated spin-wave dispersion for $J$=130 meV (red line, see below), strongly suggesting a magnetic origin of this feature.

\begin{figure}
\includegraphics[width=3.3in]{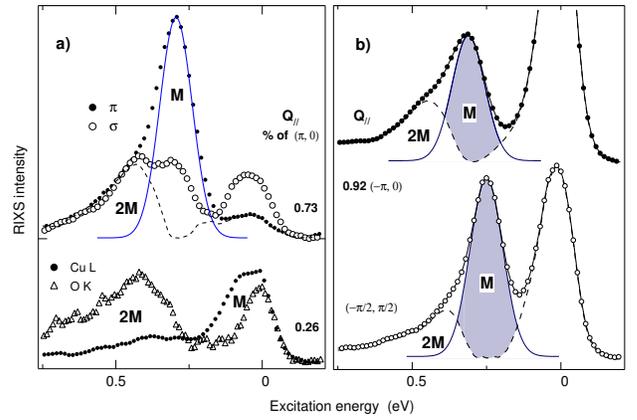}
\caption{(color online) (a) (top) RIXS spectra for $\sigma$ and $\pi$ polarization and Q$_{\|}$=0.58 \AA$^{-1}$ along (100). The dashed line is the difference between the $\pi$ spectrum and the gaussian line shape (M, solid blue line) representing the single-magnon contribution. (bottom) Comparison of Cu L-edge and O K-edge
RIXS for  Q$_\|$=0.19  \AA$^{-1}$. (b) RIXS spectra at $0.92(-\pi,0)$ and $(-\pi/2,\pi/2)$ measured with $\sigma$ polarization. The two magnons (2M) to single magnon (M) intensity ratio is 0.26 at $(-\pi/2,\pi/2)$, and increases to 0.49 at $0.92(-\pi,0)$.}
\label{fig:2}
\end{figure}

A model independent analysis was performed by fitting the main peak of the spectrum to a resolution-limited gaussian line shape representing the single-magnon (M) contribution (Fig.\ 2 top). Subtracting this line shape from the raw spectrum yields asymmetric features (dashed line) on both sides of the magnon peak. The low-energy one at $\sim$ 50 meV, contains the elastic line, phonon losses and possibly a residue due to the approximate magnon line shape. The higher-energy feature reflects two-magnon (2M) and higher-order excitations, which give rise to the continuum above the single-magnon dispersion curve in Fig.\ 1c. Details of this analysis have negligible impact on the extracted magnon energy. The typical uncertainty is $\pm$10 meV. For LCO, a similar approach yields a magnon dispersion in excellent agreement with INS \cite{Ament}.

\begin{figure}
\includegraphics[width=0.40\textwidth]{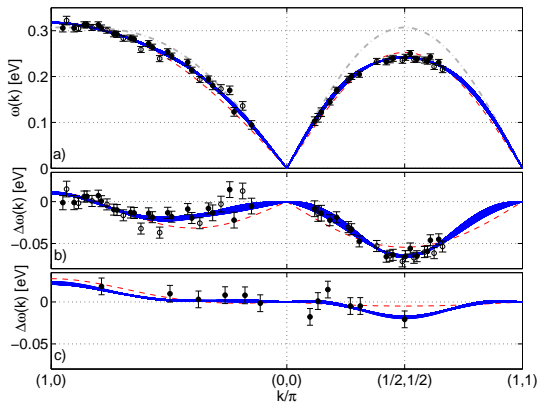}
\caption{(color online) a) Magnon energies extracted from the RIXS data. Open and closed symbols stem from 2 independent measurements on different samples. Dot-dashed line is a NN Heisenberg model with $J=130$~meV. Red dashed line is a NN Hubbard model fit for $t=0.261\pm0.004$~eV and $U=1.59\pm0.04$~eV. Blue lines are the further neighbor Hubbard fits. b) same as a) with NN Heisenberg dispersion subtracted to better visualize details of the dispersion. The blue band shows the spread in dispersions obtained for fits with $1.9$~eV$<U<4$~eV.
c) The fit to the neutron data on La$_2$CuO$_4$ \cite{Coldea}, shifted by $(\pi, \pi)$ for ease of comparison.
}
\label{fig:3}
\end{figure}

An interesting and unexplored aspect of RIXS is the possibility to separate single- and multi-magnon contributions by exploiting their different edge- and polarization-dependent cross sections, as illustrated in Fig.\ 2 (a). The top panel shows spectra (Q$_\|=0.58$~\AA$^{-1}$) for two polarizations, normalized in the 2M region for ease of comparison. The relative intensity of the single-magnon peak is reduced for $\sigma$ polarization.
The bottom panel compares Cu L$_3$ and O K edge (1$s$, 530 eV; $\Delta$E=60 meV) data, for the same Q$_\|$ (0.19  \AA$^{-1}$). Similar to the Cu K-edge case, the O K edge spectrum contains only the 2M continuum, which was suggested by previous experiments \cite{Harada,Bisogni}, and is now clearly resolved around 0.4 eV for this Q$_\|$. By contrast, the L$_3$ line shape exhibits a prominent single magnon loss at $\sim$ 0.1 eV.
Figure 2 (b) ilustrates the spectral line shapes near the zone boundary points $(-\pi/2,\pi/2)$ and $(-\pi,0)$. Near $(-\pi,0)$ the magnon peak is weaker, and spectral weight is transferred to the higher energy continuum. The comparison strongly suggests that the quantum effect observed by neutrons in CFTD \cite{Christensen} is also present in SCOC, and most likely in the cuprates in general. This observation is important for theories arguing that superconductivity occur from strong magnetic fluctuations \cite{Anderson,Lee}, although further development of the RIXS cross-section and more quantitative predictions from these theories are required before definite conclusions can be drawn.

The Q-dependence of the magnon energy extracted from the data of Fig. 1, and from similar data for the (110) direction and two different samples, is summarized in Fig. 3. The RIXS data are consistent with the small-Q results from INS, but cover for the first time the full dispersion up to the boundary of the MBZ. They reveal a striking 70 meV difference between the magnon energies of 310~meV at ($\pi$,0) and 240~meV at ($\pi/2$,$\pi/2$). This can be compared with the smaller $\sim$20 meV dispersion in LCO \cite{Coldea}. Dispersion along the zone boundary in all cuprates is also predicted by recent theory, which however underestimates the ($\pi$,0) energy in SCOC by almost 50 meV \cite{Wan}.

For the simple $S$=1/2 2D Heisenberg model with NN exchange, linear spin-wave theory predicts a constant magnon energy $\hbar\omega$=2$J$ along the MBZ boundary. First order quantum corrections uniformly renormalize the dispersion by a factor Z$_c$=1.18. Numerical results \cite{Singh95prb,Syljuasen} and neutron data on CFDT  \cite{Ronnow2001prl,Christensen} have established that the magnon energy for purely NN exchange is actually 6\% larger at ($\pi/2$,$\pi/2$) than at ($\pi$,0).
The dispersion in Fig. 3 and in LCO is in the opposite direction to this quantum effect. It could be reproduced by adding freely adjustable further-neighbor exchange interactions. However, the Heisenberg (spin-only) hamiltonian is the low-energy projection of an electronic system at half filling, and a better approach is to systematically consider higher orders to this projection. Indeed the dispersion in LCO \cite{Coldea} was described by projecting the one-band Hubbard model with effective Coulomb repulsion $U$ to 4th order in the NN hopping $t$, giving rise to further-neighbor exchange interactions $J$, $J'=J''$ and $J_c$.  The same approach gives for SCOC an unphysically low value of $U=1.59\pm0.04$~eV and $t=0.261\pm0.004$~eV. This indicates that the Hubbard model with only NN hopping has reached its limit. A more plausible approach is to include further-neighbor hoppings $t'$ and $t''$ \cite{Delannoy}. We therefore extended the analysis to 4th order in $t$, $t'$ and $t''$ and find that this approach give a more reasonable range of $U$, is consistent with higher energy RIXS and ARPES, and provide better fits to the shape of the dispersion (Fig. 3b).

\begin{figure}
\includegraphics[width=0.45\textwidth]{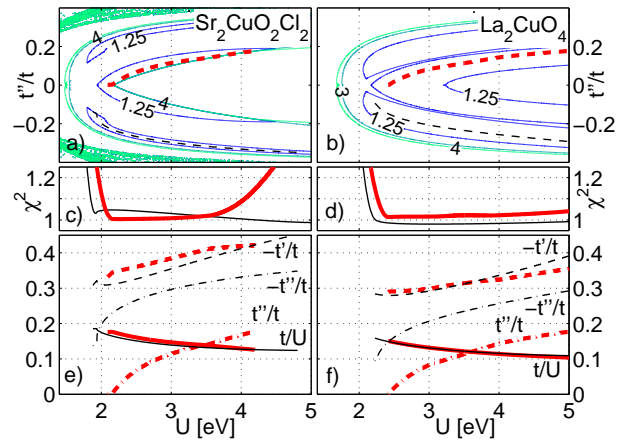}
\caption{(color online)
Results of extended Hubbard model fits for SCOC (left) and LCO (right).
a-b) $\chi^2$ after fitting $t$ and $t'$ for fixed $U$ and $t''$. Dashed lines mark the minima versus $U$, giving c-d) and e-f). c-d) $\chi^2$ versus $U$, showing the range of $U$ that our data can support. e-f) extracted hopping parameters $t$, $t'$ and $t''$ as function of $U$. For given $U$, the uncertainty on the $t$s is around 5~meV. In all panels, thick red lines are for $t''>0$ and thin black lines for $t''<0$.
}
\label{fig:4}
\end{figure}

Magnetic excitations provide accurate information on the interactions, but do not directly probe $U$. A more direct probe of the effective $U$ was found in the higher energy part of the RIXS Cu-K spectra from the sister compound Ca$_2$CuO$_2$Cl$_2$, where a 2.5-4~eV dispersive feature analyzed within the one-band Hubbard model was reported consistent with $U=3.5$~eV \cite{Hasan}. However, both the experimental and the numerical accuracy of such determinations could be improved in the future. Therefore we performed fits as a function of fixed $U$, as summarized in Fig.\ \ref{fig:4}. The dispersion is symmetric in the signs of $t'$ and $t''$, and leads to two possible solutions with $t't''<0$ and $t't''>0$ respectively. In compliance with ARPES and theoretical estimates, we assume $t'<0$, and lean towards $t''>0$. Both solutions for $t''t\lessgtr0$ constrain $U$ to larger than $\sim2$~eV and smaller than $\sim4$~eV for $t''t>0$, and give essentially the same $t/U$ and $t'/t$, which depend only weakly upon the chosen $U$, respectively from 0.17 to 0.12 and from -0.31 to -0.42.
Hence, our data provide strict constraints on the effective parameters that can be used in the one-band Hubbard model for SCOC: $U$ larger than $1.9$~eV; significant second neighbor hopping $|t'/t|>0.31$; and a unique set of hopping parameters for a given $U$ (Fig.\ \ref{fig:4}e). For $U=3.5$~eV, we obtain $t/U=0.139\pm0.004$, $t'/t=-0.41\pm0.01$ or -0.38 $\pm0.01$ and $t''/t=0.14\pm0.01$ or $-0.32\pm0.01$. These parameters are roughly consistent with ARPES results from SCOC \cite{Leung}, which were described by $U=3.5$~eV, $t=0.35$~eV, $t'/t=-0.35$ and $t''/t=0.22$, with the accuracy of the comparison likely set by the broad ARPES linewidth. Thereby, we have derived a consistent description within a single model of both spin-wave and ARPES spectra.

To gain insight into the origin of the zone-boundary dispersion, we performed the same extended Hubbard model analysis for LCO (Fig.\ \ref{fig:4} right), using the 10~K data by Coldea \emph{et al.} \cite{Coldea}. Also for LCO we find that the spin-wave dispersion can be described by a larger $U$ more compatible with higher energy probes, and that there is significant further neighbor hopping, albeit slightly weaker than in SCOC. This trend, and our quantitative hopping parameters, are consistent with recent LDA and LDA+U calculations \cite{Pavarini,Yin2009}, confirming experimentally that the main effect of the apical atom (oxygen in LSCO, chlorine in SCOC) is to influence the further-neighbor hopping terms $t'$ and $t''$.
For the projected Heisenberg hamiltonian, this means that the difference in spin-wave dispersion comes not from different main interactions $J$ and $J_c$, but from the many additional further-neighbor hopping paths and hence magnetic interactions. The existence of so-called `ring exchange' $J_c$ coupling 4 spins on a plaquette generated significant theoretical efforts to understand the consequences of this term. An important conclusion from our work is that this term is not unique and that many further interactions and larger 4-spin loops have similar weight.

In summary, we have employed RIXS to obtain qualitative and quantitative new insight into the magnetic excitation spectrum of the representative two-dimensional antiferromagnetic insulator \scoc.
Measuring the full spin-wave spectrum, we found a large zone boundary dispersion, which we could reproduce with an extended $t$-$t'$-$t''$-$U$ Hubbard model, placing quantitative constraints on the hopping parameters. Most notably, we demonstrate sizeable longer range hopping, and henceforth magnetic interactions. Taking into account the quantum corrections generated by these higher-order hopping terms is essential to achieve a quantitative description of the ground state properties of SCOC, namely the reduced value of its ordered moment \cite{Delannoy}.
In a broader perspective, these results establish an important reference and suggest a general method, requiring only small crystals, to address the nature of the ground state, and the evolution of magnetic correlations throughout the phase diagram of the cuprates.

The SAXES instrument at the ADRESS beamline of the Swiss Light Source was jointly built by Paul Scherrer Institut and Politecnico di Milano.
We gratefully acknowledge discussions with M. Mourigal, M. Gingras, J.-Y. Delannoy, F. Vernay and B. Normand, and support from the Swiss NSF and the MaNEP NCCR.

\end{document}